\documentclass[preprint,5p]{elsarticle}
\usepackage{graphicx}
\usepackage{amsmath}
\usepackage{amssymb}
\usepackage{bm}
\def\ie{{ i.e. }}
\def\eg{{ e.g. }}
\def\bra{\langle} 
\def\ket{\rangle} 
\def\R{\mathbb{R}}
 
\def\cD{\mathcal{D}}
\def\cN{\mathcal{N}}

\def\d{\partial}
\begin{document}
\begin{frontmatter}
\title{A remark on gauge invariance in wavelet-based quantum field theory}
\author[izks]{S.Albeverio}
\address[izks]{Interdisziplin\"are Zentrum f\"ur Komplexe Systeme, 
University of Bonn,  Bonn, 
D-53115, Germany} 
\ead{sergio.albeverio@yahoo.com} 
\author[jinr,iki]{M.V.Altaisky}
\address[jinr]{Joint Institute for Nuclear Research, Joliot-Curie 6, Dubna, 141980, Russia}
\address[iki]{Space Research Institute RAS, Profsoyuznaya 84/32, Moscow, 117997, 
Russia}
\ead{altaisky@mx.iki.rssi.ru}
\begin{abstract}
Wavelet transform has been attracting attention as a tool for regularization of gauge theories since the first paper of 
Federbush \cite{Federbush1995}, where the integral representation of 
the fields by means of the wavelet transform was suggested: 
$$A_{\mu}(x) =  \frac{1}{C_\psi} 
\int_{\R_+ \times\R^d} \frac{1}{a^d} g \left(\frac{x-b}{a} \right) A_{\mu a}(b) \frac{dad^db}{a},$$
with 
$A_{\mu a}(b)$ being understood as the fields $A_\mu$ measured at point $b\in \R^d$ with resolution 
$a\in\R_+$. In present paper we consider a wavelet-based theory of gauge 
fields, provide a counterpart of the gauge transform for the scale-dependent 
fields: $A_{\mu a}(x)\to A_{\mu a}(x)+\d_\mu f_a(x)$, and derive the 
Ward-Takahashi identities for them.   
\end{abstract}
\end{frontmatter}

Troubles with ultraviolet divergences taken together with the fact 
that strict localisability of quantum events is just an approximation 
that cannot be reached experimentally, stimulate the efforts to construct 
a self-consistent nonlocal field theory, at the possible lack of strict 
microcausality \cite{AE1974,Battle1999,Altaisky2010prd}. 
This is specially important for gauge field theories, 
including quantum electrodynamics and quantum chromodynamics.

In local abelian gauge field theory the local phase transformation of the 
fermionic matter fields
\begin{equation}
\psi(x) \to e^{-i e f(x)} \psi(x), \quad 
\bar\psi(x) \to e^{i e f(x)} \psi(x)
\label{gftf}
\end{equation}
(where $f$ is a real-valued gauge function of the space-time variable 
$x\equiv (x_\mu),\mu=1,\ldots,d$, $e$ is the charge of the matter fields $\psi$),
is accompanied by the substitution of space-time derivatives by covariant derivatives
$$
\frac{\d}{\d x_\mu} \equiv \d_\mu \to D_\mu \equiv \d_\mu + i e A_\mu, 
$$
which makes the theory invariant with respect to the local phase transformation \eqref{gftf}, 
if the {\em gauge field} $A_\mu$ transforms accordingly:
\begin{equation}
A_\mu \to A_\mu + \d_\mu f.
\label{gftb}
\end{equation}

The heuristic generating functional of such a theory is invariant under transformations \eqref{gftf},\eqref{gftb} if the source terms and the gauge fixing terms are invariant,
 heuristically it is thus given by 
\begin{eqnarray}
\nonumber Z[J,\bar\eta, \eta] &=& \cN \int \cD A_\mu \cD \bar\psi \cD \psi 
\exp \left(i \int L_{eff} d^dx  \right) \\
        &\equiv& \left\bra \exp \left(i \int L_{eff} d^dx \right) \right\ket 
\label{gf1} \\
\nonumber L_{eff} &=&  \imath\bar\psi \gamma^\mu D_\mu\psi -m\bar\psi\psi - \\ 
\nonumber &-&\frac{1}{4} F_{\mu\nu}F^{\mu\nu} - \frac{1}{2\alpha} (\d^\mu A_\mu)^2+ \\
&+& J^\mu A_\mu + \bar\eta \psi + \bar\psi\eta, \label{gf1l}
\end{eqnarray} 
where $L_{eff}$ is the effective Largangian,  $\cN$ is normalisation constant, $\alpha$ is a gauge fixing parameter, $\gamma$ are the gamma matrices, $J,\eta,
\bar\eta$ are test-functions for $A,\bar\psi,\psi$, respectively, $F$ is the 
curvature associated with A.
The invariance of the generating functional \eqref{gf1} under the transformations 
\eqref{gftf},\eqref{gftb} is ensured by so-called Ward-Takahashi identities 
\cite{Ward1950,Takahashi1957}.

The aim of present paper is to formulate a theory of the gauge fields 
$A_{\mu a}(x)$ that depend on both the position $x$ and the resolution $a$. 
As in previous papers 
\cite{AltSIGMA06,AltSigma2007,Altaisky2010prd} this is done by substituting 
the fields in the effective Lagrangian \eqref{gf1l} 
in terms of their continuous wavelet transforms:  
\begin{eqnarray}
A_{\mu a}(x) = \frac{1}{C_g} \int_{\R_+ \times\R^d} \frac{1}{a^d} g\left(\frac{x-b}{a}\right) A_{\mu a}(b) \frac{dad^db}{a} \label{iwt}, \\
\nonumber x,b\in\mathbb R^d, a>0,
\end{eqnarray}
where $C_g$ is a positive normalisation constant of the {\em basic wavelet} $g$,
satisfying the admissibility condition 
\begin{equation}
C_g = \int_0^\infty |\tilde g(ak)|^2\frac{da}{a}<\infty,
\label{adcf}
\end{equation}
and where $\tilde g$ means the continuous Fourier transform of $g$; see \eg \cite{Daub10,Chui1992} for reviews on the continuous wavelet transform. 

The substitution \eqref{iwt} makes the  effective Lagrangian \eqref{gf1l}  
into an effective Lagrangian for a nonlocal field theory. That is why the local gauge invariance principle 
\eqref{gftf} should be reconsidered for such a theory. Using the ideas from 
nonlocal gauge field theory \cite{Efimov1985}, we assume that the local phase invariance 
of the matter fields should be preserved  under the substitution \eqref{iwt} and the 
gauge transformations of the scale-dependent gauge fields 
$$
A_{\mu a}(b) =  \int_{\R^d} \frac{1}{a^d} \bar g\left(\frac{x-b}{a}\right) A_\mu(x) d^dx
$$
(where $\bar g$ is the complex conjugation of the basic wavelet $g$)
should be chosen accordingly  to keep that invariance. This implies the transformation conditions:
\begin{eqnarray}
\nonumber \psi(x) & \to & \psi(x) e^{\frac{-ie}{C_g}\int_{\R^d\times \R_+} \frac{1}{a^d} g\left(\frac{x-b}{a}\right) f_a(b) \frac{dad^db}{a} },    \\
A_{\mu a}(x) & \to & A_{\mu a} + \frac{\partial f_a(x)}{\partial x_\mu} \label{gft-b},
\end{eqnarray}
where 
$$f_a(b) =  \int_{\R^d} \frac{1}{a^d} \bar g\left(\frac{x-b}{a}\right) f(x) d^dx$$ 
is the continuous wavelet transform of the gauge function $f$ of the original 
local theory \eqref{gftb}. 
In the infinitesimal form (\ie up to order two in the power expansion of 
the exponent) this leads to the transformation law 
$$
\psi(x)  \to  \psi(x)\left[1 -\imath \frac{e}{C_g}\int \frac{1}{a^d} g\left(\frac{x-b}{a}\right) f_a(b) \frac{dad^db}{a}\right] .
$$
Because of the linearity of the wavelet transform the equation \eqref{gft-b} guarantees 
the gauge transform \eqref{gftb} for ordinary local gauge  
fields. 

Let us now specify the gauge theory and the Ward-Takahashi identities 
\cite{Ward1950} for the theory 
of scale-dependent fields $A_{\mu a}$.

The effective Lagrangian itself is gauge invariant by construction and only the source term acquires 
a multiplication by the factor 
$$
\exp \left(i \int \left[-\frac{1}{\alpha} (\d^\mu A_\mu)\d^2 f + J^\mu \d_\mu f 
-i e f (\bar\eta \psi - \bar\psi \eta) \right] d^dx\right),
$$
which can approximately (up to the second order term in a heuristic expansion 
in a power series)  be  represented by a first order term for ``small f'', that is 
$1+i\delta$ with 
$$
\delta\equiv \int d^dx \left[-\frac{1}{\alpha}\d^2 (\d^\mu A_\mu)  -\d^\mu  J_\mu 
-\imath e (\bar\eta \psi - \bar\psi \eta) \right] f(x).
$$
Let us replace the fields in the equation above by their integral representations in terms 
of wavelet transform \eqref{iwt}. 

Integrating by parts we put the Laplacian $\sum_\mu \frac{\d^2}{\d x_\mu^2}\equiv \d^2$ onto the gauge fixing parameter $f$. Heuristically:
\begin{eqnarray}
\nonumber \delta &=& \Bigl\langle \int d^dx \Bigl[
-\frac{1}{\alpha} \frac{1}{C_g^2} \int \frac{1}{a_1^d}
g^\mu \left(\frac{x-b_1}{a_1}\right) A_{\mu a_1}(b_1)  \times \\
\nonumber &\times & d\mu(a_1,b_1)\d^2_{x^2} 
\int \frac{1}{a_2^d} g\left(\frac{x-b_2}{a_2}\right) f_{a_2}(b_2) \times \\ 
\nonumber &\times& d\mu(a_2,b_2) \Bigr]  
-\frac{1}{C_g^2} 
\int \left[\d^\mu J_{\mu a}(b)\right]f_a(b)d\mu(a,b) -  \\ 
\nonumber &-&\frac{i e}{C_g^3} \int\left[ \bar\eta_{a_1}(b_1)\psi_{a_3}(b_3) -\bar\psi_{a_1}(b_1)\eta_{a_3}(b_3) 
\right]  \times \\ 
\nonumber &\times& f_{a_2}(b_2) \frac{1}{(a_1a_2a_3)^d} \bar g \left(\frac{x-b_1}{a_1} \right) 
g\left(\frac{x-b_2}{a_2} \right) \times \\
\nonumber &\times& g\left(\frac{x-b_3}{a_3} \right) d^dxd\mu(a_1,b_1)d\mu(a_2,b_2)d\mu(a_3,b_3) \\
& &\label{mfe}
\Bigr\rangle,
\end{eqnarray}
where $g^\mu \equiv \frac{\d g}{\d x_\mu}$ is the gradient of the basic wavelet function, $d \mu(a,b)= \frac{da d^d b}{a}$ can be looked upon as the measure on affine 
group \cite{Carey1976,DM1976} written in $L^1$ norm \cite{HM1998}, and the curly brackets 
$\bra \ldots \ket$ denote the functional averaging mean value obtained by Feynman functional integration \eqref{gf1}.
Introducing the matrix elements of operators between wavelet basic functions
\begin{eqnarray*}
\nonumber T(1,2) \equiv \int  g\left(\frac{x-b_1}{a_1}\right)
\d^2 g\left(\frac{x-b_2}{a_2}\right) \frac{d^dx}{(a_1a_2)^d}, \\
\nonumber T^{(\mu)}(1,2) \equiv \int  g^\mu\left(\frac{x-b_1}{a_1}\right)\d^2 g\left(\frac{x-b_2}{a_2}\right)\frac{d^dx}{(a_1a_2)^d}    \\
M(1,2,3) \equiv  \\ \nonumber
\equiv \int  \bar g \left(\frac{x-b_1}{a_1} \right)g\left(\frac{x-b_2}{a_2} \right) g\left(\frac{x-b_3}{a_3} \right) \frac{d^dx}{(a_1a_2a_3)^d} , 
\end{eqnarray*}
we heuristically derive the Ward-Takahashi identities for the scale-dependent fields $A_{\mu a}$.
In terms of the above defined operators $T,T^{(\mu)},M$  the variation term \eqref{mfe} can be written in the form 
\begin{eqnarray*}
\nonumber \delta &=& \Bigl\langle \int -\frac{1}{\alpha} \frac{1}{C_g^2} T(1,2) \d^\mu_{b_1} A_{\mu a_1}(b_1) 
f_{a_2}(b_2)d\mu(a_1,b_1)  \\
\nonumber &\times& d\mu(a_2,b_2) - \frac{1}{C_g^2}\d^\mu J_{\mu a_2}(b_2)d\mu(a_2,b_2) \\
\nonumber &-&\frac{\imath e}{C_g^3} \left[{
\bar\eta_{a_1}(b_1)\psi_{a_3}(b_3) -\bar\psi_{a_1}(b_1)\eta_{a_3}(b_3) 
}\right]  \\
&\times& f_{a_2}(b_2)M(1,2,3) d\mu(a_1,b_1) d\mu(a_2,b_2) d\mu(a_3,b_3)
\Bigr\rangle.
\end{eqnarray*} 
To obtain the heuristic variation of the generating functional the fields 
should be substituted by corresponding functional derivatives:
$$
\psi \to \frac{1}{\imath} \frac{\delta}{\delta\bar\eta}, \quad 
\bar \psi \to \frac{1}{\imath} \frac{\delta}{\delta\eta}, \quad 
A \to \frac{1}{\imath} \frac{\delta}{\delta J},
$$
with all variations taken with respect to the measure on the affine 
group $d\mu(a,b)$.

Assuming that the full variation of the generating functional with 
respect to gauge transformations is zero, this gives the functional 
equation
\begin{eqnarray*}
\Bigl[ \frac{i}{\alpha C_g^2} T(1,2) \d^\mu_1 
\frac{\delta}{\delta J^\mu(1)} - \frac{1}{C_g^2} \d^\mu J_{\mu}(2) 
-\frac{e}{C_g^3} 
\bigl(\bar \eta(1) \frac{\delta}{\delta\bar\eta(3)} - \\
-\eta(3)\frac{\delta}{\delta\eta(1)}
\bigr) 
M(1,2,3) d\mu(a_1,b_1)d\mu(a_3,b_3) 
\Bigr]Z[\bar\eta,\eta,J]=0.
\end{eqnarray*}
To  derive the Ward-Takahashi equations for connected Green functions we 
heuristically substitute 
$$ Z=\exp(i W). $$ This gives the following equation in functional derivatives
\begin{eqnarray*}
-\frac{i}{\alpha} \frac{1}{C_g^2} \int d\mu(a_1,b_1) 
\left[\d^\mu_1 \frac{\delta W}{\delta J^\mu(1)} \right] T(1,2) - \\
- \frac{1}{C_g^2} \d^\mu J_{\mu}(2) 
- \frac{i e}{C_g^3} \int \left(\bar \eta(1) \frac{\delta W}{\delta\bar\eta(3)} 
-\eta(3)\frac{\delta W}{\delta\eta(1)}
 \right) \times \\
\times M(1,2,3) d\mu(a_1,b_1)d\mu(a_3,b_3) = 0.
\end{eqnarray*}
To get the equations for the vertex functions $\Gamma[\psi,\bar\psi, A_\mu]$ we apply 
the functional Legendre transform 
$$
\Gamma[\psi,\bar\psi, A_\mu] = W[\eta,\bar\eta,J] - \int \bar\eta \psi + \bar\psi\eta + JA
$$
to the latter equations.
Doing so we arrive heuristically to the following equation in functional derivatives for the 
vertex function
\begin{eqnarray}
&-&\frac{1}{\alpha} \frac{1}{C_g^2}\d^\mu_1 A_\mu(1)  T(1,2) 
+\frac{1}{C_g^2} \d^\mu_{2} \frac{\delta\Gamma}{\delta A^\mu(2)} \label{wt0}\\  
\nonumber &-& \frac{i e}{C_g^3} \left(\psi(3)\frac{\delta\Gamma}{\delta\psi(1)}
-\bar\psi(1)\frac{\delta\Gamma}{\delta\bar\psi(3)}
 \right) M(1,2,3)  = 0,
\end{eqnarray}
where the short-hand notation $A_\mu(1)\equiv A_{\mu a_1}(b_1),\psi(3)\equiv \psi_{a_3}(b_3)$  is used and the integration over all repeated indices is assumed. 
The Ward-Takahashi equations are heuristically derived by taking the second derivatives of the equation 
\eqref{wt0} at zero fields ($A=\psi=\bar\psi=0$). This gives
\begin{eqnarray*}
& &\frac{1}{C_g^2}\d^\mu_2\frac{\delta^3\Gamma[0]}{\delta\bar\psi(x_1)\delta\psi(y_1)\delta A^\mu(2)}
-\frac{i e}{C_g^3} \frac{\delta^2}{\delta\bar\psi(x_1)\delta\psi(y_1)} \\
\nonumber &\times& \left( \psi(3)\frac{\delta\Gamma}{\delta\psi(1)} -\bar\psi(1)\frac{\delta\Gamma}{\delta\bar\psi(3)}\right)M(1,2,3)  = 0.
\end{eqnarray*}

Performing the heuristic functional differentiation and using the symmetry under 
the permutation $2\leftrightarrow3$ after 
the integration we have the Ward-Takahashi identities
\begin{eqnarray}
\nonumber &-& \frac{1}{C_g^2}\d^\mu_x \frac{\delta^3\Gamma[0]}{\delta\bar\psi(x_1)\delta\psi(y_1)\delta A^\mu(x)} \\ 
&=& i e \frac{\delta^2\Gamma[0]}{\delta\bar\psi(1)\delta\psi(y_1)}M(x_1,x,1)- \label{wt8} \\ 
\nonumber &-&i e \frac{\delta^2\Gamma[0]}{\delta\bar\psi(x_1)\delta\psi(1)}M(1,x,y_1),
\end{eqnarray}
where $x,x_1,y_1$ are arbitrary position-resolution arguments 
$x\equiv(a_x,b_x)\in \R_+\times \R^d$.
Following ,\eg, \cite{Ryder1985} we define vertex functions and inverse propagators in the Fourier space
\begin{eqnarray}
\nonumber & &\int d^db_x d^db_{x_1}d^db_{y_1}\exp\left(i(p'b_{x_1}-pb_{y_1}-q b_x)\right) \times \\
\nonumber &\times& \frac{\delta^3\Gamma[0]}{\delta\bar\psi(x_1)\delta\psi(y_1)\delta A^\mu(x)}  \\
&=& i e (2\pi)^d \delta(p'-p-q) \Gamma_{\mu a_{x_1} a_{y_1} a_x}(p,q,p'),
\label{v3} \\
\nonumber & &\int d^db_{x_1}d^db_{y_1}\exp\left(\imath(p'b_{x_1}-pb_{y_1})\right) \frac{\delta^2\Gamma[0]}{\delta\bar\psi(x_1)\delta\psi(y_1)}  \\
&=& \imath (2\pi)^d \delta(p'-p) {S'}_{a_{x_1}a_{y_1}}^{-1}(p). \label{v2} 
\end{eqnarray}
Using the definitions (\ref{v3},\ref{v2}) we multiply equation \eqref{wt8} by 
$e^{\imath(p'b_{x_1}-pb_{y_1}-q b_x)}$ and integrate over $d^db_x d^db_{x_1}d^db_{y_1}$.
This gives
\begin{eqnarray}
q^\mu \Gamma_{\mu a_4 a_3 a_1}(p,q,p+q) =  \label{wtes}\\
\nonumber = \int \frac{da_2}{a_2}
S_{a_1 a_2}^{-1}(p+q) \tilde M_{a_2a_3a_4}(p+q,q,p)\\
\nonumber - \int \frac{da_2}{a_2} \tilde M_{a_1a_3a_2}(p+q,q,p) S_{a_2 a_4}^{-1}(p), 
\end{eqnarray}
where 
\begin{eqnarray*}
\tilde M_{a_1 a_2 a_3}(k_1,k_2,k_3) &=& (2\pi)^d \delta^d(k_1-k_2-k_3) \times \\
&\times& \overline{\tilde g}(a_1k_1)\tilde g(a_2 k_2) \tilde g(a_3 k_3)
\end{eqnarray*}
is the Fourier image of the vertex operator $M$.

The equation \eqref{wtes} is a non-local analog of the ordinary (local) 
Ward-Takahashi equation in Fourier space.

\section*{Acknowledgement}
The paper was supported  by DFG Project 436 RUS 113/951.

\end{document}